\documentclass[notitlepage, preprint, nofootinbib]{revtex4-1}

\usepackage[utf8]{inputenc} 
\usepackage{xcolor}
\usepackage{amsmath}
\usepackage{amsfonts, amssymb}
\usepackage{graphicx}
\graphicspath{ {./images/} }
\usepackage{float}
\usepackage{caption}
\usepackage{subcaption}
\usepackage{natbib}
\usepackage{hyperref}

\begin{document}

\title{More on the relativistic images produced in gravitational lensing}

\author{Angelika Ghazale$^{1,2}$}
\email{ag224mu@student.lnu.se}
\affiliation{$^1$ Department of Physics and Electrical Engineering, IFE, Linnaeus University, Universitetsplatsen 1, 352 52 Kalmar, Sweden,\\
$^2$ Dipartimento di Scienza e Alta Tecnologia, Universit\`a degli Studi dell'Insubria, via Valleggio 11, I-22100 Como, Italy}

\author{Oliver F. Piattella$^{1,2,3,4}$}
\affiliation{$^1$ Dipartimento di Scienza e Alta Tecnologia, Universit\`a degli Studi dell'Insubria, via Valleggio 11, I-22100 Como, Italy,\\
$^2$ INFN Sez. di Milano, Via Celoria 16, 20126, Milano\\
$^3$ N\'ucleo Cosmo-ufes, Universidade Federal do Esp\'irito Santo, avenida F. Ferrari 514, 29075-910 Vit\'oria, Esp\'irito Santo, Brazil
\\
$^4$ Como Lake centre for AstroPhysics (CLAP), DiSAT, Università dell’Insubria, via Valleggio 11, 22100 Como, Italy}

\begin{abstract}
    We investigate the gravitational lensing properties and the formation of relativistic images associated with black holes modeled by the Reissner–Nordström and Kerr space-time geometries. In particular, we perform numerical computations of the deflection angles, image angular positions, magnification factors (including demagnification), and time delays between distinct images, and we systematically quantify the dependence of these observables on the electric charge and spin parameters of the black hole.    
\end{abstract}

\maketitle

\newpage

\section{Introduction}

Over the past several decades, the theoretical investigation of strong gravitational lensing by compact astrophysical objects, in particular black holes, has become a major focus of research. The notion of gravitational deflection of light traces back to the pioneering work of Einstein \cite{einstein_lens-like_1936} and its first empirical confirmation during the 1919 Eddington expedition \cite{lang_123_1920}. Nonetheless, it was not until the latter half of the twentieth century that gravitational lensing emerged as a well-established and mature subfield of astrophysics and cosmology, supported by both rigorous theoretical developments and systematic observational studies \cite{refsdal_possibility_1964, blandford_fermats_1986}.

A particularly extreme case of strong gravitational lensing occurs when light rays experience deflection angles exceeding \(2\pi\), such that photons execute one or multiple complete orbits around the black hole in the vicinity of the photon sphere before reaching the observer. This phenomenon gives rise to two infinite sequences of images formed near the photon sphere, commonly termed relativistic images, a designation introduced by Virbhadra in 2000 \cite{virbhadra_schwarzschild_2000}. Whereas the formation of multiple images in conventional strong lensing is a well-established observational phenomenon, the specific relativistic images arising from such light loops remain a purely theoretical prediction of general relativity and have not yet been directly detected.

In 2001, Bozza \cite{bozza_strong_2001} developed an analytical framework for computing large deflection angles in gravitational lensing, known as the strong deflection limit (SDL), which exhibits a logarithmic divergence as light rays approach the photon sphere of a black hole. This method has subsequently been applied to a variety of spacetime metrics and within different gravitational theories to analyze higher-order relativistic images formed in the vicinity of black holes \cite{eiroa_reissner-nordstrom_2002, eiroa_gravitational_2011, eiroa_gravitational_2012, bozza_quasi-equatorial_2003, bozza_time_2004, bozza_comparison_2008, bozza_gravitational_2002, ghosh_analytical_2022, filho_antisymmetric_2024, tsukamoto_gravitational_2021, tsukamoto_retrolensing_2022}.

On the observational side, gravitational lensing can cause background sources to manifest as multiple images, elongated arcs, or nearly complete rings, commonly referred to as Einstein rings. The first confirmed case of strong gravitational lensing was reported in 1979, when the quasar $Q0957+561$ was detected as two distinct images exhibiting nearly identical redshifts and spectral characteristics, attributable to the gravitational potential of an intervening galaxy \cite{walsh_0957_1979}. Subsequently, in 1987, the first Einstein ring was identified using the Very Large Array (VLA) network of radio telescopes, providing striking and direct visual confirmation of the gravitational lensing phenomenon \cite{hewitt_unusual_1988}.

Beyond its original role as an observational test of general relativity, gravitational lensing has developed into a highly versatile and powerful probe in contemporary cosmology. It enables detailed measurements of mass distributions in galaxies and galaxy clusters, provides stringent constraints on the properties and distribution of dark matter and on the nature of dark energy, facilitates the investigation of distant and otherwise observationally inaccessible astrophysical sources, and offers an independent method for determining the Hubble constant through time-delay cosmography \cite{schneider_gravitational_1992, perlick_gravitational_2004, bartelmann_gravitational_2010}.

In this article, we present the analytical framework underlying the strong-field limit of gravitational lensing, together with a comprehensive numerical investigation carried out for the Schwarzschild, Reissner–Nordström (RN), and Kerr spacetimes. For the numerical computations, we adopt as reference the physical parameters of the supermassive black hole Sagittarius A$^*$ (Sgr A$^*$). Specifically, we use the following values \cite{virbhadra_schwarzschild_2000}: mass $m = 2.8 \times 10^6 M_\odot$, distance to the observer $D_d = 8.5$ kpc, and mass-to-distance ratio $m/D_d \approx 1.57 \times 10^{-11}$. The source position is chosen such that $D_{ds}/D_s = 1/2$, and its angular location with respect to the optical axis is taken to be $\beta = \pm 0.075$ radians $\approx \pm 4.29718^{\circ}$ \cite{virbhadra_schwarzschild_2000},\footnote{The use of the $\pm$ symbol in the source position indicates that the source may lie on either side of the optical axis, corresponding to two angular configurations that are symmetric with respect to the axis.} unless stated otherwise.

Although the dimensionless spin parameter of Sgr A$^*$ has been estimated to be approximately $a/2m = 0.9 \pm 0.06$ \cite{daly_new_2023}, a broader range of spin values is examined in order to systematically assess the role of spin in Kerr spacetime. Furthermore, the impact of electric charge is investigated by modeling Sgr A$^*$ as a non-rotating black hole, specifically within the framework of the Reissner–Nordström spacetime.

We compute the angular positions, deflection angles, magnification factors, strong-field limit coefficients, lensing observables, and time delays of relativistic images up to second order in the winding number $(n=2)$. These results yield testable predictions for forthcoming very-long-baseline interferometry (VLBI) measurements and quantitatively characterize the impact of both charge and spin on the associated lensing observables.

\section{Lensing Profile or Relativistic Images}

\begin{figure}[ht]
  \centering
  \includegraphics[width=0.9\textwidth]{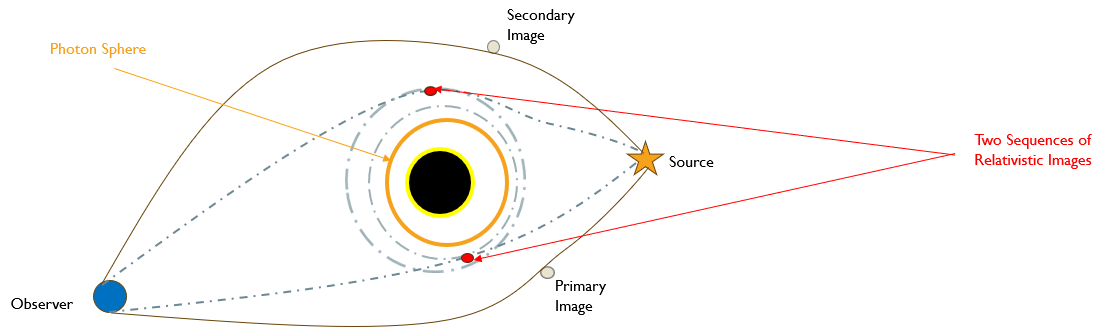}
  \caption{\label{fig: relativistic images} Illustration showing the formation of the two relativistic image sequences due to the looping around the photon sphere of the black hole. This also conveys the location of the primary and secondary images obtained.}
\end{figure}

Relativistic images are formed when light rays pass very close to the photon sphere and loop around the black hole's photon sphere once or several times in both clockwise and counterclockwise directions before finally reaching the observer. See Fig. \ref{fig: relativistic images}.

Relativistic images appear very close to the black hole, near the photon sphere, and are extremely faint because only a small fraction of the light from the source passes close enough to form these images. Theoretically, light can loop around the black hole an infinite number of times, leading to an infinite sequence of relativistic images on each side. Each image corresponds to light that has completed an additional loop around the black hole, with an additional image created closer to the shadow.

\begin{figure}[ht]
  \centering
  \includegraphics[width=0.45\textwidth]{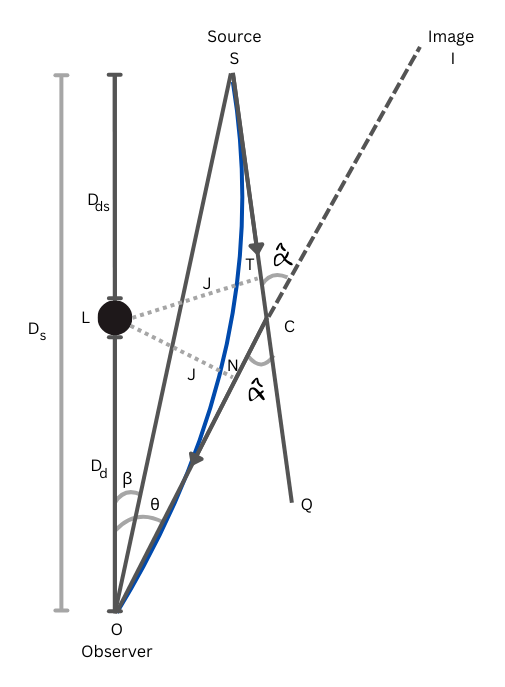}
  \caption{\label{3} Lens diagram with lens $L$, observer $O$, source $S$ and image $I$. Distances $D_{ds}$, $D_d$ and $D_s$ are the respective distances from the lens to source, observer to lens and observer to source. $OL$ is the optical axis in which the angular positions of the source $\beta$ and the image $\theta$ are measured; $\hat{\alpha}$ is the deflection angle of the light ray (in blue). $SQ$ and $OI$ are tangents to the light ray (in blue) at the $S$ and $O$ respectively. $LN$ and $LT$ are perpendiculars to $OI$ and $SQ$ respectively which are the impact parameter $J$. This figure is an adaptation to the figure presented in \cite{virbhadra_schwarzschild_2000}.}
\end{figure}

With reference to the lensing system in Fig. \ref{3}, the lens equation and the impact parameter $J$ can be expressed as \cite{virbhadra_schwarzschild_2000}: 
\begin{equation}\label{lens}
\tan \beta = \tan \theta -  \frac{D_{ds}}{D_s}[\tan\theta + \tan (\hat{\alpha} - \theta)]\,, \qquad J=D_d\sin\theta\,,
\end{equation}
where the deflection angle in the strong deflection limit (SDL) of the light ray is the following \cite{bozza_strong_2001}:
\begin{equation}\label{deflection}
    \hat{\alpha}(\theta)=-\Bar{a}\log\left(\frac{\theta D_d}{J_p}-1\right)+\Bar{b}+O(J-J_p)\,, 
\end{equation}
where $\Bar{a}$ and $\Bar{b}$ are the so-called strong limit coefficients and depend only on the metric components; $J_p$ is the critical impact parameter evaluated at the radius of the photon sphere.

\begin{figure}[ht]
  \centering
  \subfloat[\label{1}Diagram showing the different images formed. $I_p$ and $I_s$ are the primary and secondary images respectively. $I_r$ are the relativistic images formed by the clockwise and counter-clockwise winding. In blue are the trajectories of the light rays.]{\includegraphics[width=0.4\textwidth]{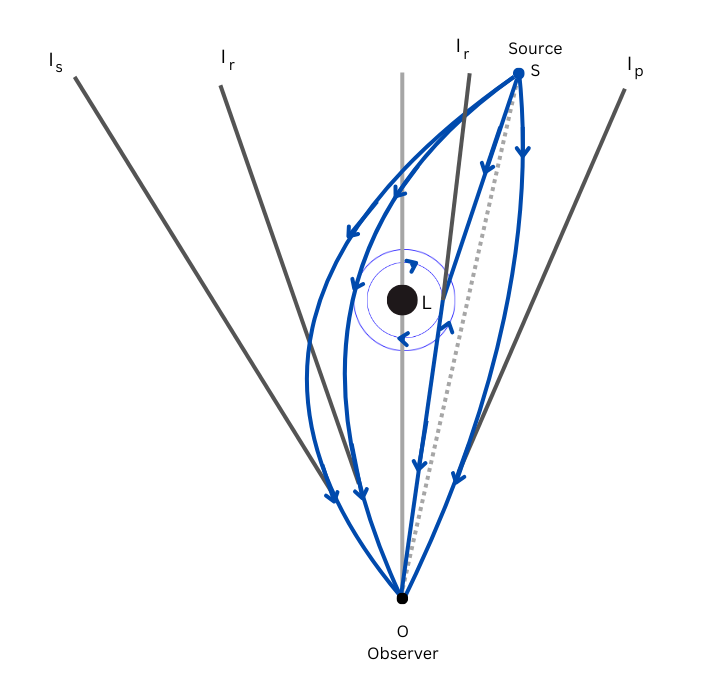}\label{fig:f1}}
  \hfill
  \subfloat[\label{2} Diagram Showing the Primary Einstein ring (in blue) and the infinite sequence of relativistic Einstein rings (in gray) obtained in the case of perfect alignment.]{\includegraphics[width=0.44\textwidth]{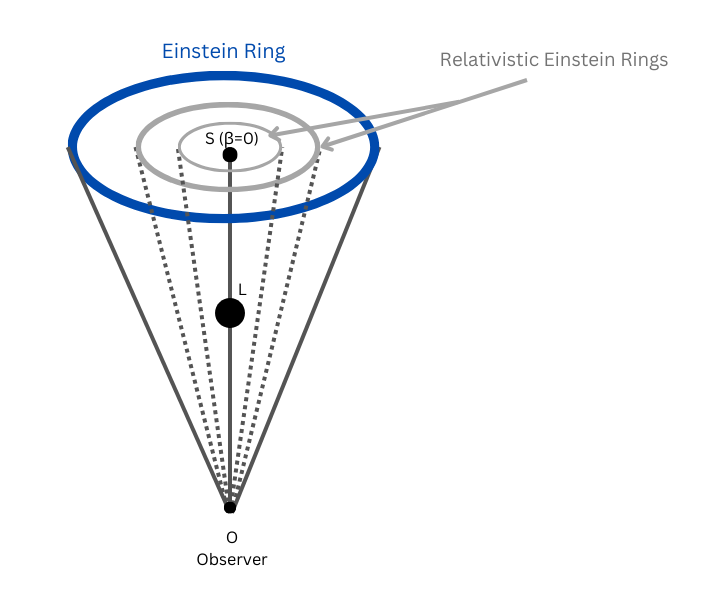}\label{fig:f2}}
  \caption{The relativistic images.}
\end{figure}

In Fig. \ref{1}, the lens, observer, and source are slightly misaligned, thereby generating two infinite sequences of relativistic images, each found at an angular position :
\begin{equation}
    \theta_n=\theta^0_n+\frac{D_s}{D_{ds}D_d}\frac{J_pe_n}{\Bar{a}}(\beta-\theta^0_n)\,, \qquad \theta^0_n=\frac{J_p}{D_d}(1+e_n)\,, \qquad e_n=e^\frac{\Bar{b}-2n\pi}{\Bar{a}}\,,
\end{equation}
where $n$ is the order of the image.

When the source is aligned with the lens and the observer $(\beta=0)$, as in Fig. \ref{2}, one expects the formation of an infinite sequence of Einstein rings, with angular positions:
\begin{equation}
\label{eq:relateinstring}
    \theta_n^E=\left(1-\frac{D_s}{D_{ds}D_d}\frac{J_pe_n}{\Bar{a}}\right)\theta^0_n\,.
\end{equation}
The magnification of these relativistic images can be found as follows:
\begin{equation}\label{mag}
    \mu_n=\left(\frac{\beta}{\theta}\frac{d\beta}{d\theta}\right)^{-1}\bigg|_{\theta^0_n}=\frac{1}{\beta}\left[\frac{D_s}{D_{ds}D_d}\frac{J_pe_n}{\Bar{a}}\theta^0_n\right]=\frac{J_p^2D_se_n(1+e_n)}{\Bar{a}\beta D_{ds}D_d^2}\,.
\end{equation}
The magnification $\mu_n$ decreases very rapidly with $n$ and is proportional to $(J_p/D_d)^2$. Hence, it is useful to consider the case where the outermost (first order) image $\theta_1$ is resolved as a single image, and all the other images are grouped together in a limiting value: 
\begin{equation}
    \theta_\infty = \frac{J_p}{D_d}\,.
\end{equation}
This represents the asymptotic position of the higher-order relativistic images in the limit $n \rightarrow \infty$, expressed in terms of the minimum impact parameter $J_p$.

\begin{figure}[ht]
  \centering
  \includegraphics[width=0.4\textwidth]{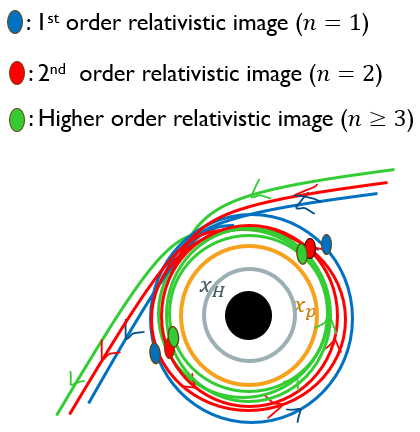}
  \caption{\label{10} Illustration showing the looping of light rays and the formation of the higher order relativistic images with their positions around the photon sphere of the black hole.}
\end{figure}

In this approximation, the angular separation between the first relativistic image and the other images is:
\begin{equation}
\label{eq:s}
    s = \theta_1 - \theta_\infty = \theta_\infty e^{\frac{\Bar{b}-2\pi}{\Bar{a}}}\,.
\end{equation}
The ratio of the flux between the first relativistic image and the sum of the fluxes for all the other relativistic images, except for the first, is given by \cite{bozza_gravitational_2002}: 
\begin{equation}
    r = e^{\frac{2 \pi}{\Bar{a}}}\,,
\end{equation}
which, converted in magnitude, is:
\begin{equation}
    r_{\rm mag} = \frac{5\pi}{\Bar{a} \ln 10}\,.
\end{equation}

\begin{figure}[ht]
  \centering
  \includegraphics[width=1.0\textwidth]{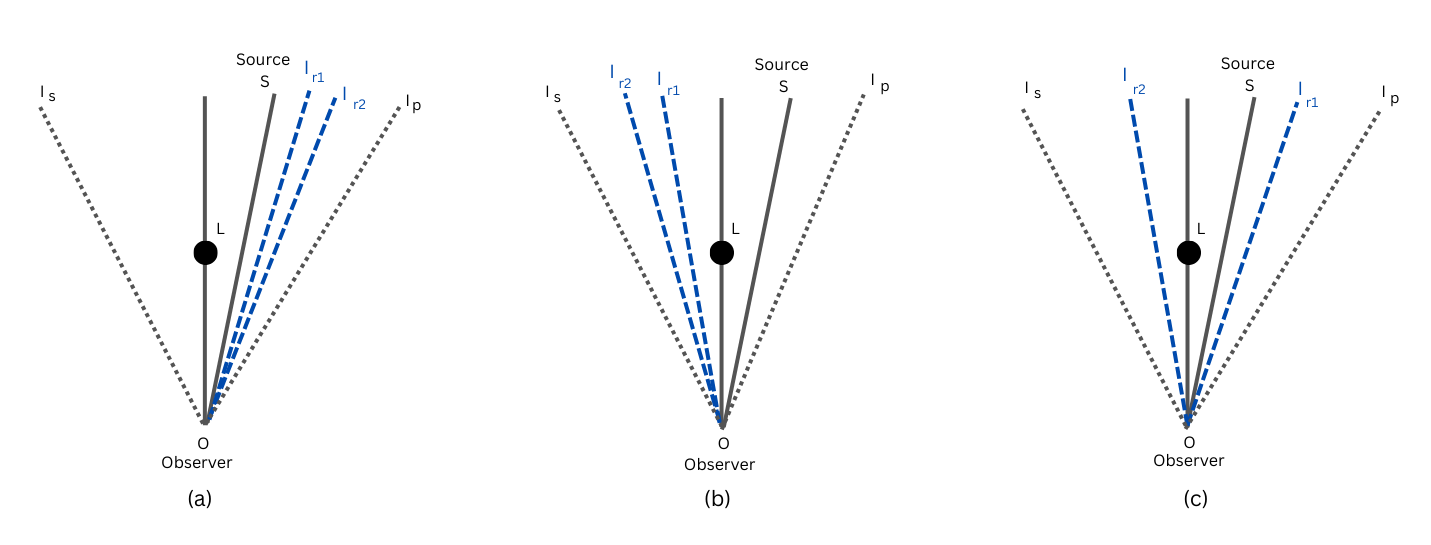}
  \caption{\label{4}Illustration showing the different relativistic image locations. On each side of the lens $L$ is an infinite series of relativistic images formed. $I_{r1}$ and $I_{r2}$ stand for the first and second order relativistic images. $I_p$ and $I_s$ are the primary and secondary images respectively.}
\end{figure}

Since relativistic images are formed very close to the photon sphere, a similar approach to that of the deflection angle was followed to develop a logarithmic time delay for the photon winding in terms of the impact parameter by \cite{bozza_time_2004}:
\begin{equation}
\label{eq:approxtime}
    \Tilde{T}(J)=-\Tilde{a}\log\left(\frac{J}{J_p}-1\right)+\Tilde{b}+O(J-J_p)\,,
\end{equation}
where $\Tilde{a}$ and $\Tilde{b}$ are new coefficients that depend on the black hole's parameters contained in the geodesic equation $dt/dx$.

When calculating the time delay of the relativistic images, it is important to determine their location with respect to the lens. The formation of relativistic images is obtained when the photons wind around the black hole in both the clockwise and the anticlockwise directions. Thus, they can appear on both sides of the lens or on either of them. This is illustrated in Fig. \ref{4}.

In case (a), the two images are on the same side of the lens and the source. In case (b), the two images are on the same side of the lens but opposite to the source. For both cases (a) and (b), their time delay is given by $\triangle T^s_{n,l}$. In case (c), the two images are on opposite sides; therefore, their time delay corresponds to $\triangle T^o_{n,l}$. Here, both $n$ and $l$ are the winding numbers of the ray for the obtained image. After identifying the position of the relativistic images, the final form of the time delay in the equation will be equal to $\triangle T^s_{n,l}$ or $\triangle T^o_{n,l}$ for images on the same or opposite sides of the lens, respectively. These will be derived for the respective spacetime geometry in sections \ref{Reissner-Nordstrom Spacetime} and \ref{Kerr_Spacetime}.

\section{Reissner-Nordström Spacetime}\label{Reissner-Nordstrom Spacetime}

The line element of the isotropic black hole with charge $\epsilon$ is \cite{reissner_uber_1916}:
\begin{equation}
    ds^2 = \left(1-\frac{2m}{r}+\frac{\epsilon^2}{r^2}\right)dt^2-\left(1-\frac{2m}{r}+\frac{\epsilon^2}{r^2}\right)^{-1}dr^2-r^2(d\theta^2+\sin^2\theta d\phi^2)\,.
\end{equation}
The radius of the outer horizon and the photon sphere radius are: 
\begin{equation}
    r_+=m+(m^2-\epsilon^2)^\frac{1}{2}\,, \qquad r_p=\frac{3}{2}m\left(1+\sqrt{1-\frac{8\epsilon^2}{9m^2}}\right)\,.
\end{equation}
For later numerical calculations, it is useful to define dimensionless distances and charge: 
\begin{equation}
    x=\frac{r}{2m}, \qquad x_0=\frac{r_0}{2m}, \qquad q=\frac{\epsilon}{2m}\,.
\end{equation}

\begin{figure}[ht]
     \centering
     \begin{subfigure}[b]{0.6\textwidth}
         \centering
         \includegraphics[width=\textwidth]{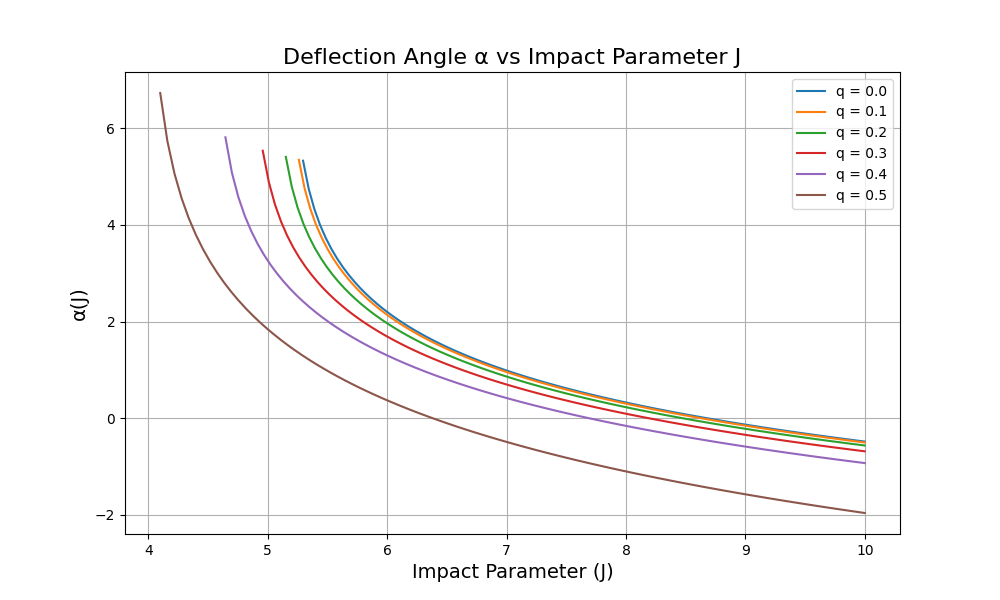}
         \caption{$\hat{\alpha}$ (rad) vs $J$}
         \label{7}
     \end{subfigure}
     \hfill
     \begin{subfigure}[b]{0.6\textwidth}
         \centering
         \includegraphics[width=\textwidth]{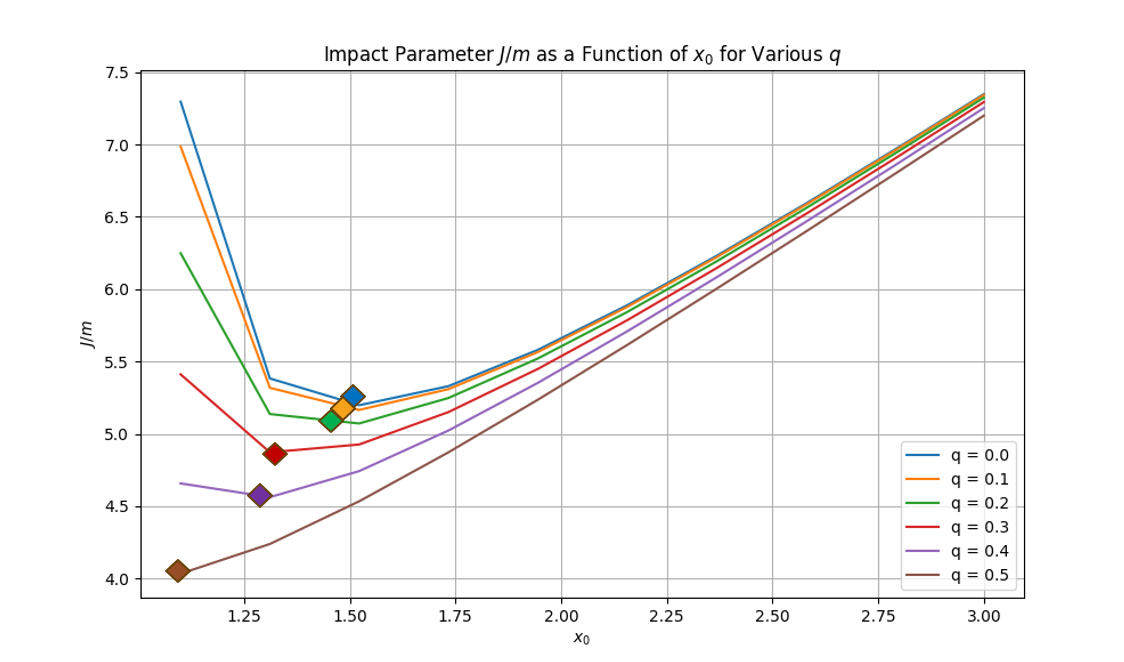}
         \caption{$J/m$ vs $x_0$}
         \label{8}
     \end{subfigure}
        \caption{In panel (a) we display the plots of the deflection angle $\hat{\alpha}$ (in radians) vs the impact parameter $J$. Panel (b) shows the dimensionless impact parameter $J/m$ vs the distance of closest approach $x_0$. Both graphs are obtained for different charge $q$ values varying from $q=0$ (the Schwarzschild geometry, depicted in blue) to the extremal case $q = 1/2$ (as $\epsilon = m$).}
        \label{fig:1}
\end{figure}

\begin{table}[ht]
    \centering
    \begin{tabular}{|c|c|c|c|c|c|c|c|c|c|c|}
        \hline
        $\epsilon$ & $q$ & $x_H$ & $x_p$ & $J_p$ & $\Bar{a}$ & $\Bar{b}$ & $\hat{\alpha}_1 (\mu as)$ & $\theta_1 (\mu as)$ & $\hat{\alpha}_2 (\mu as)$ & $\theta_2 (\mu as)$ \\ \hline
        0 & 0 & 1 & 1.5 & 5.196$m$ & 1 & -0.4 & $2\pi+33.69$ & 16.848 & $4\pi+33.65$ & 16.826 \\ \hline
        0.2$m$ & 0.1 & 0.989 & 1.486 & 5.161$m$ & 1.0045 & -0.399 & $2\pi+33.47$ & 16.736 & $4\pi+33.42$ & 16.713 \\ \hline
        0.4$m$ & 0.2 & 0.958 & 1.444 & 5.052$m$ & 1.0197 & -0.397 & $2\pi+32.77$ & 16.387 & $4\pi+32.72$ & 16.363 \\ \hline
        0.6$m$ & 0.3 & 0.9 & 1.368 & 4.858$m$ & 1.052 & -0.396 & $2\pi+31.52$ & 15.762 & $4\pi+31.46$ & 15.734 \\ \hline
        0.8$m$ & 0.4 & 0.8 & 1.242 & 4.545$m$ & 1.1231 & -0.414 & $2\pi+29.52$ & 14.760 & $4\pi+29.44$ & 14.721 \\ \hline
        1$m$ & 0.5 & 0.5 & 1 & 4$m$ & 1.4142 & -0.733 & $2\pi+26.09$ & 13.046 & $4\pi+25.90$ & 12.954 \\ \hline
    \end{tabular}
    \caption{Characteristic quantities for the lensing process considered in this paper; the photon sphere radius $x_p$, critical impact parameter $J_p$ and strong limit coefficients $\Bar{a}$ and $\Bar{b}$ for different values of the charge $q$. Values for deflection angles $\hat{\alpha}_1$ and $\hat{\alpha}_2$ and angular positions $\theta_1$ and $\theta_2$ for the first (outermost) and second (innermost) order relativistic images are also shown for the same set of charge $q$ values. All angles are given in microarcseconds $(\mu as)$.}
    \label{tab:1}
\end{table}

\begin{figure}[ht]
  \centering
  \includegraphics[width=0.6\textwidth]{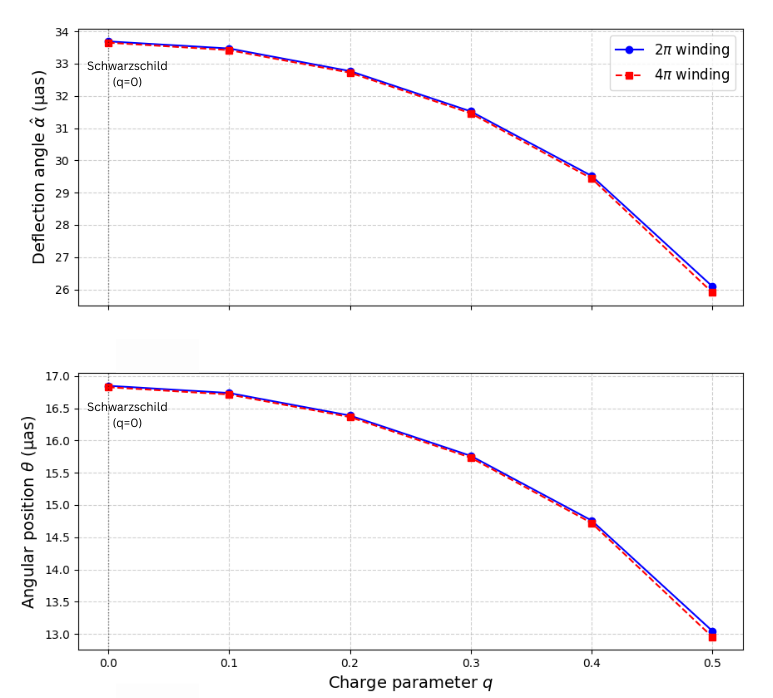}
  \caption{\label{9}Plot for the deflection angle $\hat{\alpha}$ and angular position $\theta$ presented in Tab. \ref{tab:1} vs the charge $q$ for the innermost and outermost images.}
\end{figure}

The characteristic quantities for the strong field gravitational lensing by a charged black hole, computed for a range of dimensionless charge values from $q=0$ (Schwarzschild) to $q=0.5$ (extremal), are summarized in Tab.\ref{tab:1} and Fig.\ref{fig:1} visually complement this data.

The overarching trend observed is a weakening of the lensing effect as the black hole's charge $q$ increases. This behavior is a direct consequence of the change in spacetime geometry described by the RN metric. The inclusion of charge introduces a repulsive gravitational component that partially counteracts the attractive force due to mass. Consequently, for a given mass $m$, a charged black hole constitutes a less compact gravitational lens. The photon sphere moves inward because the gravitational field is weaker, allowing light to orbit closer to the event horizon without being captured. The reduction in $J_p$ indicates a smaller black hole shadow, and the decrease in $\hat{\alpha}$ and $\theta$ signifies that the relativistic images formed are located closer to the optical axis.

The strong deflection limit coefficients $\bar{a}$ and $\bar{b}$ also exhibit a clear dependence on $q$. The increase in $\bar{a}$ indicates that the divergence of the deflection angle near the photon sphere becomes sharper with higher charges. The evolution of $\bar{b}$ towards more negative values further characterizes the precise functional form of this divergence.

Although the photon orbit $x_p$ decreases with increasing $q$, the deflection angle $\hat{\alpha}$ also decreases. For instance, in the extremal case ($q=0.5$), the light comprising the relativistic images approaches as close as $x_p=1.0$, significantly closer than $x_p=1.5$ in the Schwarzschild case. Despite this closer approach, the deflection is weaker ($2\pi+26.09 \mu as$ vs $2\pi+33.69 \mu as$).

This is resolved by recognizing that the total deflection is not a local function of the minimum radius $x_0$ alone, but rather a global integral of the spacetime curvature along the entire null geodesic. Although the path for the higher $q$ case is closer to the black hole, the spacetime curvature at all comparable radii is weaker due to the charge's repulsive contribution. This phenomenon is clearly illustrated in Fig. 6 (b): for any fixed distance of closest approach $x_0>x_p$, the corresponding impact parameter $J$ is smaller for the Schwarzschild case than for any of the charged cases with the same $x_0$. 

\begin{figure}[ht]
     \centering
     \begin{subfigure}[b]{0.45\textwidth}
         \centering
         \includegraphics[width=\textwidth]{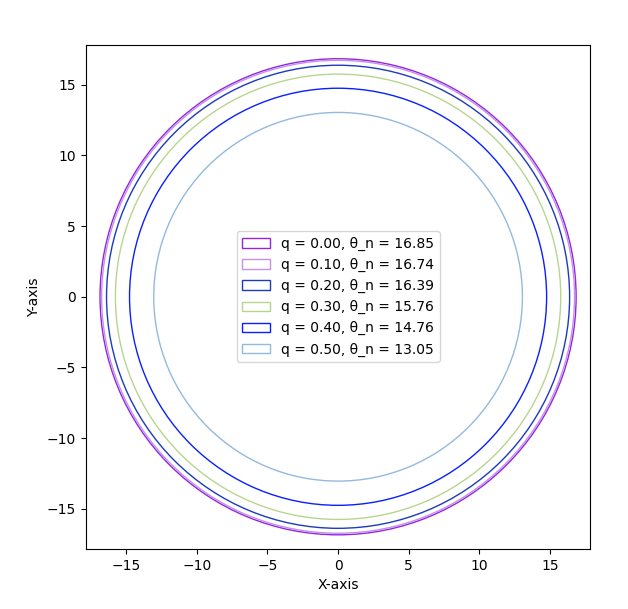}
         \caption{$\theta_1^E (\mu as)$}
         \label{777}
     \end{subfigure}
     \hfill
     \begin{subfigure}[b]{0.45\textwidth}
         \centering
         \includegraphics[width=\textwidth]{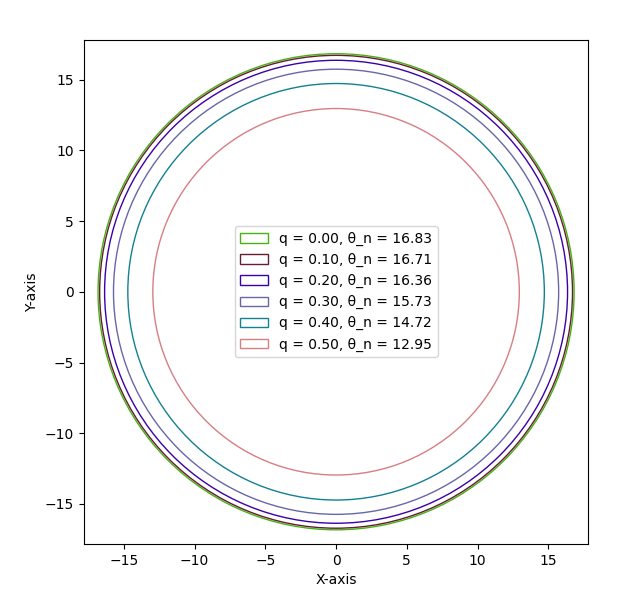}
         \caption{$\theta_2^E (\mu as)$}
         \label{888}
     \end{subfigure}
        \caption{Plots for the outermost Einstein rings $\theta_1^E$ (left) and the innermost relativistic Einstein rings $\theta_2^E$ (right) vs the charge $q$ in perfect alignment ($\beta=0$). $q=0$ represents the Schwarzschild Einstein rings.}
        \label{fig:99}
\end{figure}

The angular positions of the relativistic images manifest themselves as Einstein rings in the case of perfect alignment between the source, lens, and observer ($\beta=0$). Fig. \ref{fig:99} displays the angular radii of the outermost ($\theta^E_1$) and innermost of the considered second-order ($\theta^E_2$) Einstein rings as a function of the black hole charge $q$.

The results confirm that the weakening of the lensing effect with increasing charge, already established, directly translates to the properties of the rings as well. As $q$ increases from $0$ to $0.5$, both $\theta^E_1$ and $\theta^E_2$ shrink towards the optical axis, decreasing by approximately $22.5\%$ and $23.1\%$, respectively. 
\begin{figure}[ht]
  \centering
  \includegraphics[width=0.6\textwidth]{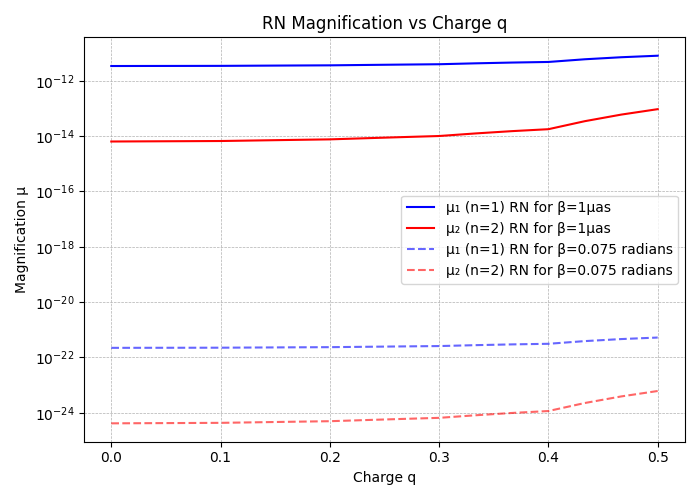}
  \caption{\label{11}Magnifications $\mu_n$ for the first and second order images for the source with angular position $\beta=1\mu as$ and $\beta=0.075$rad.}
\end{figure}

The observability of these relativistic images is determined by their magnification. Fig. \ref{11} illustrates the magnifications $\mu_n$ of the first and second-order images for two source positions: a nearly perfectly aligned case ($\beta=1 \mu as$) and a significantly misaligned case ($\beta=0.075 rad$).

The most striking feature is the extreme dependence of the magnification on the source alignment. For the near-aligned configuration, the magnifications are larger by more than ten orders of magnitude compared to the misaligned case. This is a feature of gravitational lensing; magnification is dramatically enhanced when the source, lens, and observer are closely aligned.

For a fixed source position, the magnification exhibits a mild but consistent increase with black hole charge. This trend is more pronounced for the second-order image and in the misaligned case. Although a higher charge leads to a weaker deflection angle and smaller image positions, it results in stronger magnifications. This behavior can be understood through the lens equation. The magnification is inversely proportional to the derivative of the deflection angle with respect to the position of the image (\ref{lens} and \ref{mag}). The increasing values of the strong field coefficient $\bar{a}$, which signifies a sharper divergence of the deflection angle, likely lead to a higher rate of change in the lens mapping, thereby increasing the magnification for higher-charge black holes despite the weaker overall gravitational pull.

\begin{table}[ht]
\centering
  \begin{tabular}{|l|p{2cm}|p{2cm}|p{2cm}|p{2cm}|p{2cm}|p{2cm}|l|}
    \hline
    \textbf{q} & \textbf{0}  & \textbf{0.1}  & \textbf{0.2}  & \textbf{0.3}  & \textbf{0.4}  & \textbf{0.5} \\
    \hline  $\theta_1 (\mu as) $ & 16.848 & 16.736 & 16.387 & 15.762 & 14.760 & 13.046\\
    \hline  $\theta_\infty (\mu as)$ & 16.826 & 16.713 & 16.363 & 15.734 & 14.721 & 12.953 \\
    \hline  $s (\mu as)=\theta_1 - \theta_\infty $ & 0.02149 & 0.0228 & 0.0239 & 0.0278 & 0.0387 & 0.0925 \\
    \hline  $r$  & 535.491 & 520.628 & 474.280 & 392.974 & 268.943 & 85.023 \\
    \hline  $r_{mag}$ & 6.8218 & 6.7913 & 6.6900 & 6.4859 & 6.0741 & 4.8238 \\
    \hline 
    \end{tabular}
    \caption{\label{tab:table5.8}Numerical calculation of the remaining inner packed images $\theta_\infty$ which represents the asymptotic position approached by a set of images in the limit $ n\rightarrow \infty$, the separation $s$ between the first relativistic image and the packed images, $r$ the flux of the relativistic image and $r_{mag}$ which is the difference between the magnitude of the first image and all other images for different charge values $q$.}
\end{table}

Although all image positions $\theta$ decrease with respect to charge $q$, the separation between consecutive image positions given in Tab.\ref{tab:table5.8}, increases with charge. This indicates that the higher-order images are affected more strongly by the change in spacetime geometry than the outermost image. The increase in separation $s$ is a direct consequence of the changing strong field deflection coefficients $\bar{a}$ and $\bar{b}$, which govern the logarithmic divergence of the deflection angle \ref{deflection} and thus the spacing between subsequent images.

The flux ratio $r$ and the magnitude difference $r_{mag}$ (Tab.\ref{tab:table5.8}), provide further insight into the brightness distribution among the images. The monotonic decrease of both $r$ and $r_{mag}$ with increasing $q$ indicates that the contrast in brightness between the outermost relativistic image and the inner packed images diminishes. In other words, for a highly charged black hole, the inner relativistic images are relatively brighter compared to the first image than they are for a Schwarzschild black hole. This reinforces the conclusion that charge not only changes the scale but also the relative distribution of brightness within the relativistic image family.

The time delay for relativistic images on the same side in a spherically symmetric spacetime is \cite{bozza_time_2004}:
\begin{equation}\label{eqn: deltaTs}
    \triangle T^s_{n,l}=2\pi(n-l)\frac{\Tilde{a}}{\Bar{a}}+2\sqrt{\frac{B_pJ_p}{A_p}}\left[e^\frac{\Bar{b}-2l\pi\pm\beta}{2\Bar{a}}-e^\frac{\Bar{b}-2n\pi \pm\beta}{2\Bar{a}}\right]\approx 2\pi(n-l) J_p\,.
\end{equation}
The time delay for two relativistic images on the opposite side of the lens \cite{bozza_time_2004}:
\begin{equation}
\label{eqn: deltaTo}
    \triangle T^o_{n,l}=[2\pi(n-l)-2\beta] \frac{\Tilde{a}}{\Bar{a}} +2\sqrt{\frac{B_pJ_p}{A_p}}\left[e^\frac{\Bar{b}-2l\pi-\beta}{2\Bar{a}}-e^\frac{\Bar{b}-2n\pi+\beta}{2\Bar{a}}\right] \approx [2\pi(n-l)-2\beta]J_p \,.
\end{equation}

\begin{table}[H]
    \centering
    \begin{tabular}{|c|c|c|c|}
        \hline
        q & $\triangle T^s_{2,1}$ (min) & $\triangle T^o_{2,1}$ (min) & $\triangle T^o_{1,1}$ (sec) \\ \hline
        0   & 7.5   & 7.414   & -11.396   \\ \hline
        0.1   & 7.45   & 7.366   & -11.329   \\ \hline
        0.2   & 7.296  & 7.219  & -11.122  \\ \hline
        0.3  & 7.016  & 6.958  & -10.756  \\ \hline
        0.4  & 6.57  & 6.589  & -10.182  \\ \hline
        0.5  & 5.783  & 5.893  & -9.237  \\ \hline
    \end{tabular}
    \caption{Time delay calculations for the first and second order images on the same $\triangle T^s_{2,1}$ (min) and opposite $\triangle T^o_{2,1}$ (min) side of the lens. Time delays for the first-order image on opposite sides of the lens $\triangle T^o_{1,1}$ (sec). The negative sign indicates that the signal (or light ray) arrives earlier than the direct, undeflected path. $q=0$ corresponds to the Schwarzschild BH.}
    \label{tab:2}
\end{table}

The dominant trend is a clear decrease in all time delays with increasing charge $q$. This observed decrease is a direct consequence of the reduction in the critical impact parameter $J_p$ with charge (Tab.\ref{tab:1}). The approximate forms of equations \ref{eqn: deltaTs} and \ref{eqn: deltaTo}, $\Delta T \propto J_p$, show that the time delay scales linearly with the characteristic impact parameter. As the photon sphere contracts ($x_p$ decreases) and the black hole's effective gravitational reach diminishes ($J_p$ decreases), the difference in the path lengths traveled by light rays forming consecutive relativistic images is reduced. Consequently, the temporal separation between their arrivals at the observer also shortens.

A particularly intriguing result is the behavior of $\Delta T^o_{1,1}$. This delay is negative and increases from $-11.4$ to $-9.2$ seconds with increasing charge. A negative value indicates that the signal from one of the images arrives earlier than the signal from the direct, undeflected path would. The image on one side follows a trajectory that, while highly bent, is geometrically shorter than the direct path. The trend towards a less negative value with charge indicates that this effect becomes less pronounced in more highly charged black holes, as the overall weakening of the gravitational field reduces the efficiency of these extreme light-bending shortcuts. This provides a unique temporal signature that could, in principle, distinguish between black holes with different charges.

\section{Kerr Spacetime}\label{Kerr_Spacetime}

The Kerr black hole line element in the Boyer-Lindquist coordinates is of the form: \cite{kerr_gravitational_1963}: 
\begin{equation}
\label{12}
    ds^{2}= \frac{\triangle}{\rho^{2}}(dt - a\sin^{2}\theta d\phi)^{2} - \frac{\sin^{2}\theta}{\rho^{2}}[(r^{2} +a^{2})d\phi - adt]^{2} - \frac{\rho^{2}}{\triangle}dr^{2} - \rho^{2}d\theta^{2}\,,
\end{equation}
where \begin{math} \rho^{2}=r^{2}+a^{2}\cos^{2}\theta\end{math}, \begin{math} \triangle=r^{2}+a^{2}-2mr\end{math}, $m$ is the mass and $a=l/m$ is the angular momentum per unit mass. The horizons are found to be: \begin{equation}
   r_{H\pm}=m\pm (m^{2}-a^{2})^\frac{1}{2}\,.
\end{equation} 
The surfaces of infinite redshift are found at a radius $r_S$:
\begin{equation}
\label{15}
    r_{S\pm} = m \pm (m^{2}-a^{2}\cos^2 \theta)^\frac{1}{2}\,.
\end{equation} 
The study of the photon trajectories on the Kerr spacetime will be limited to the equatorial plane, where we assume both the observer and the source to lie. For $\theta=\pi/2$, the line element \eqref{12} reduces to:
\begin{equation}
\label{13}
ds^{2}=\left(1-\frac{2m}{r}\right)dt^{2} - \frac{r^{2}}{\triangle}dr^{2} - \left(r^{2}+a^{2}+\frac{2ma^{2}}{r}\right)d\phi^{2} - \frac{4ma}{r}dtd\phi\,,
\end{equation}
whereas $r_{S\pm} = 0, 2m$.



Introducing the non-dimensional quantities:
\begin{equation}
    x=\frac{r}{2m}, \qquad x_0=\frac{r_0}{2m}, \qquad a=\frac{a}{2m}\,,
\end{equation}
the horizon, surfaces of infinite redshift and impact parameter become:
\begin{equation}
    x_{H+} = \frac{1}{2} + \left(\frac{1}{4} - a^2\right)^\frac{1}{2}\,, \qquad x_{S\pm} = 0, 1\,, \qquad \label{eq:impact}
    J = \frac{2mx_0}{(x_0-1)}\left[(x_0^2 -x_0 +a^2)^\frac{1}{2} - \frac{a}{x_0}\right]\,.
\end{equation}
The photon sphere radius is given by:
\begin{equation}
    x_p = 1+\cos\left[\frac{2}{3} \arccos{(\pm 2|a|)}\right]\,.
\end{equation}

\begin{figure}[ht]
     \centering
     \begin{subfigure}[b]{0.6\textwidth}
         \centering
         \includegraphics[width=\textwidth]{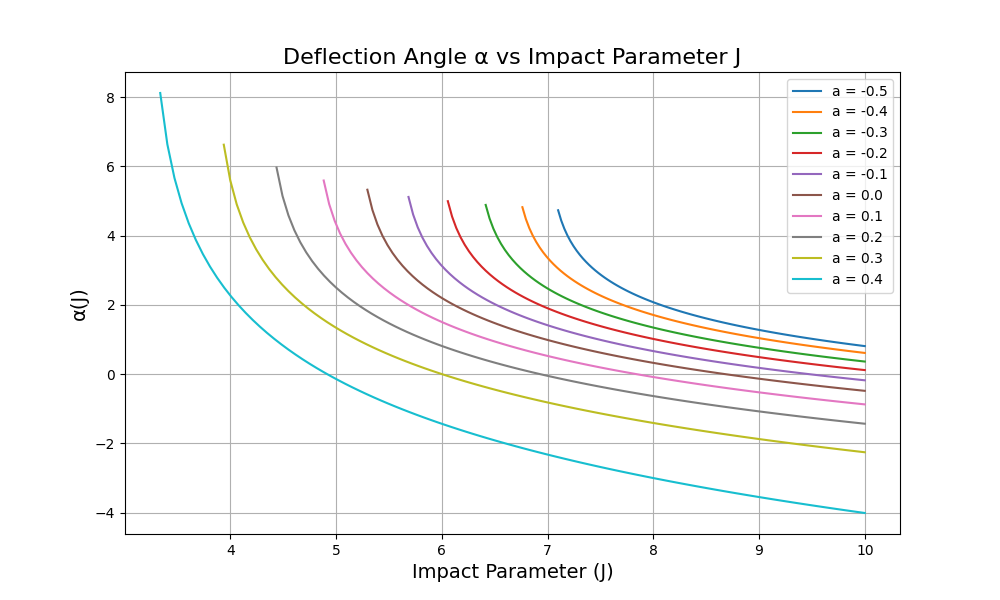}
         \caption{$\hat{\alpha}$ (rad) vs $J$}
         \label{77}
     \end{subfigure}
     \hfill
     \begin{subfigure}[b]{0.6\textwidth}
         \centering
         \includegraphics[width=\textwidth]{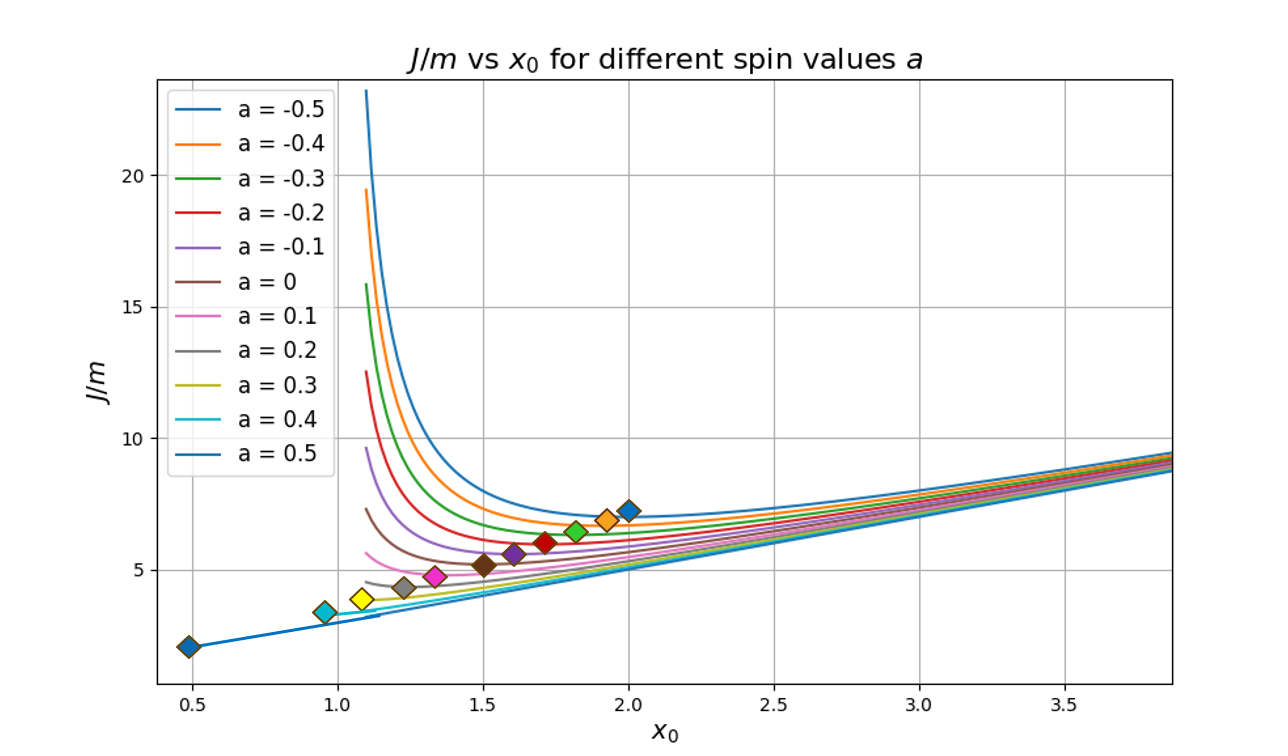}
         \caption{$J/m$ vs $x_0$}
         \label{88}
     \end{subfigure}
        \caption{(a) Plots of the deflection angle $\hat{\alpha}$(rad) \textit{vs.} the impact parameter $J$. (b) Plots of the dimensionless impact parameter $J/m$ \textit{vs.} the distance of closest approach $x_0$. Both graphs are obtained for different spin $a$ values varying from $a=0$ (Schwarzschild) to maximal $a=a/2m=0.5$ in the prograde and retrograde cases. The brown curve represents the Schwarzschild case $a = 0$.}
        \label{fig:9}
\end{figure}

\begin{table}[ht]
    \centering
    \begin{tabular}{|c|c|c|c|c|c|c|c|c|c|c|c|c|}
        \hline
        $ a (dim)$ & $a$ & $x_H$ & $x_p$ & $J_p$ & $\Bar{a}$ & $\Bar{b}$ & $\hat{\alpha}_1 (\mu as)$ & $\theta_1 (\mu as)$ & $\hat{\alpha}_2 (\mu as)$ & $\theta_2 (\mu as)$ \\ \hline
        -1m & -0.5 & 0.5 & 2 & 7m & 0.769 & -0.110 & $2\pi+45.34$ & 22.674 & $4\pi+45.33$ & 22.668 \\ \hline
        -0.8m & -0.4 & 0.8 & 1.909 & 6.662m & 0.800 & -0.143 & $2\pi+43.16$ & 21.581 & $4\pi+43.14$ & 21.573 \\ \hline
        -0.6m & -0.3 & 0.9 & 1.814 & 6.315m & 0.835 & -0.207 & $2\pi+40.91$ & 20.459 & $4\pi+40.90$ & 20.450 \\ \hline
        -0.4m & -0.2 & 0.958 & 1.715 & 5.957m & 0.878 & -0.263 & $2\pi+38.60$ & 19.302 & $4\pi+38.58$ & 19.290 \\ \hline
        -0.2m & -0.1 & 0.989 & 1.611 & 5.585m & 0.932 & -0.338 & $2\pi+36.20$ & 18.100 & $4\pi+36.17$ & 18.080 \\ \hline
        0 & 0 & 1 & 1.5 & 5.196m & 1 & -0.400 & $2\pi+33.69$ & 16.848 & $4\pi+33.65$ & 16.826 \\ \hline
        0.2m & 0.1 & 0.989 & 1.379 & 4.783m & 1.089 & -0.482 & $2\pi+31.04$ & 15.521 & $4\pi+30.97$ & 15.489 \\ \hline
        0.4m & 0.2 & 0.958 & 1.245 & 4.337m & 1.220 & -0.624 & $2\pi+28.19$ & 14.095 & $4\pi+28.09$ & 14.045 \\ \hline
        0.6m & 0.3 & 0.9 & 1.094 & 3.838m & 1.437 & -0.893 & $2\pi+25.03$ & 12.519 & $4\pi+24.84$ & 12.429 \\ \hline
        0.8m & 0.4 & 0.8 & 0.905 & 3.237m & 1.918 & -1.259 & $2\pi+21.38$ & 10.692 & $4\pi+20.98$ & 10.490 \\ \hline
        1m & 0.5 & 0.5 & 0.5 & 2m & - & - & - & - & - & - \\ \hline
    \end{tabular}
    \caption{Values corresponding to the lensing characteristics; the photon sphere radius $x_p$, critical impact parameter $J_p$ and strong limit coefficients $\Bar{a}$ and $\Bar{b}$ for different spin $a$ values. Values for deflection angles $\hat{\alpha}_1$ and $\hat{\alpha}_2$ and angular positions $\theta_1$ and $\theta_2$ for the first (outermost) and second (innermost) order relativistic images are also shown for the same set of spin $a$ values.}
    \label{tab:9}
\end{table}

\begin{figure}[ht]
  \centering
  \includegraphics[width=0.55\textwidth]{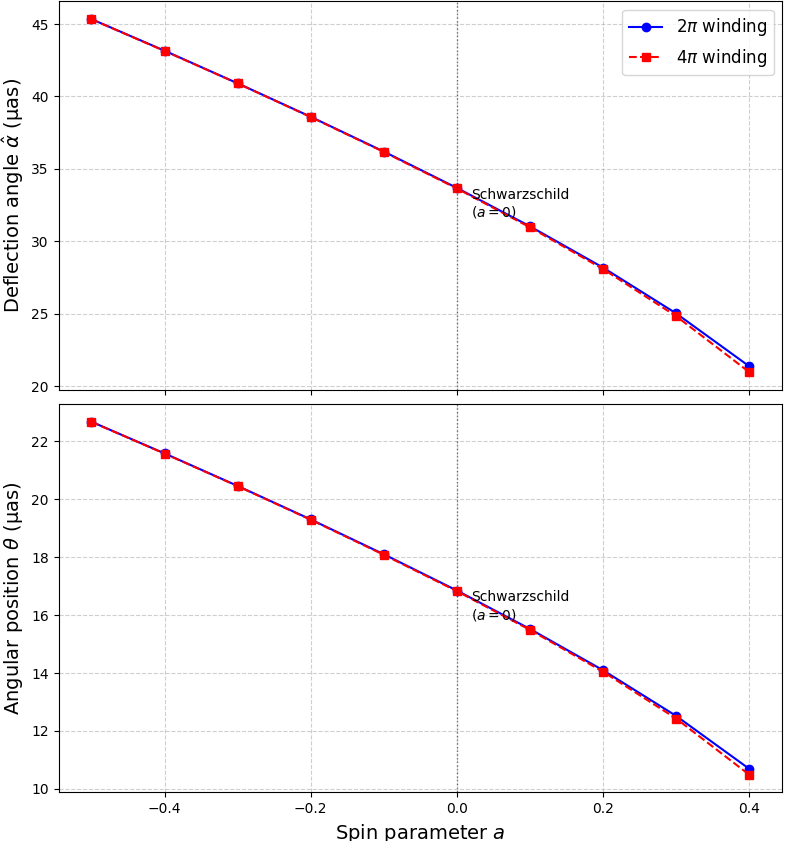}
  \caption{\label{99}Plot for the deflection angle $\hat{\alpha}$ and angular position $\theta$ presented in Tab. \ref{tab:1} \textit{vs.} the spin $a$ for the innermost and outermost images.}
\end{figure}

The gravitational lensing characteristics for a rotating Kerr black hole are summarized in Tab. \ref{tab:9} and illustrated in Fig. \ref{fig:9}. The analysis spans from the nonrotating Schwarzschild limit ($a = 0$) to rapidly rotating cases with dimensionless spin parameter $a = a/2m = \pm 0.5$, clearly distinguishing between prograde (corotating with the spin, $a > 0$) and retrograde (counterrotating, $a<0$) photon orbits.

The most pronounced effect of rotation is the breaking of spherical symmetry, leading to a stark difference between prograde and retrograde motion. This is fundamentally due to frame-dragging, an effect of GR where a rotating mass drags the surrounding spacetime along with it. For prograde orbits, the photon is carried along by the rotating spacetime, effectively reducing the centripetal force required for a circular orbit. This allows the photon sphere to move closer to the event horizon. Conversely, for retrograde orbits, the photon must fight against the dragged spacetime, requiring it to orbit at a larger distance to avoid falling in.

This is confirmed by the data, the photon sphere radius $x_p$ decreases from $1.5$ for $a=0$ to a minimum of $0.5$ for $a=0.5$ in the prograde case. For retrograde motion, $x_p$ increases to a maximum of $2$ at $a=-0.5$. A similar critical trend is observed for the impact parameter $J_p$, which decreases to $2m$ for prograde orbits and increases to $7m$ for retrograde orbits. This asymmetry directly translates into an asymmetric black hole shadow \cite{bardeen_rotating_1972}, which would appear distorted, with the prograde side appearing closer to the center and the retrograde side being stretched away.

The deflection angles, shown in Fig. \ref{99}, are strongly influenced by this asymmetry. For a fixed impact parameter $J$, retrograde orbits experience a significantly larger deflection angle than prograde orbits. This is because a retrograde photon, orbiting at a larger radius, samples a region of spacetime where the frame-dragging effect acts as an additional gravitational pull against its motion, enhancing the bending. The first-order deflection angle $\hat{\alpha}_1$ reflects this, reaching a maximum of $2\pi+45.34 \mu as$ for the extreme retrograde case ($a=-0.5$) and a minimum of $2\pi+21.38 \mu as$ for a prograde spin of $a=0.4$.

The strong field limit coefficients $\bar{a}$ and $\bar{b}$ also exhibit a clear dependence on spin. The coefficient $\bar{a}$, which governs the divergence rate of the deflection, increases significantly for prograde orbits (to $1.918$ at $a=0.4$), indicating a much sharper ``cliff'' near the photon sphere. The coefficient $\bar{b}$ becomes more negative for prograde spin and less negative for retrograde spin, further characterizing the precise logarithmic divergence for each case.

The angular positions of the relativistic images, $\theta_1$ and $\theta_2$, are direct observational consequences of the spin-induced asymmetry. For prograde orbits, the images shift closer to the optical axis, with $\theta_1$ decreasing from $16.848 \mu as $ to $10.692\mu as $. For retrograde orbits, the images move outward, with $\theta_1$ increasing to $22.674\mu as$.

For the extremal case $a=0.5$, the lensing parameters are not computed in Tab. \ref{tab:9} , as the photon orbit coincides with the event horizon in this limit, and the strong-field expansion used likely breaks down, which is why the entries are left blank.

\begin{figure}[ht]
  \centering
  \includegraphics[width=0.8\textwidth]{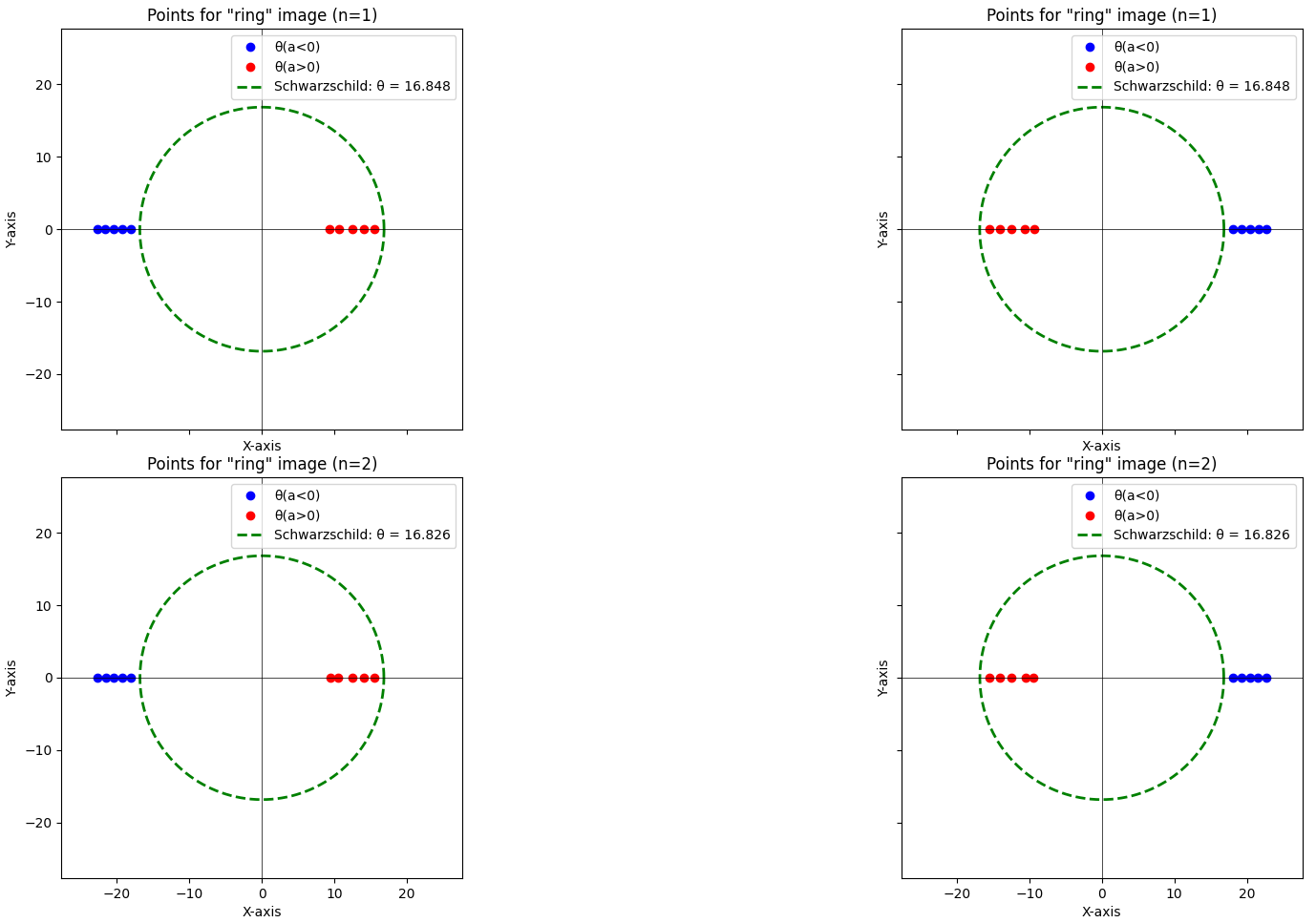}
  \caption{\label{98}Plots showing the angular positions $\theta_1^E$ and $\theta_2^E$ of the images for perfect alignment in which a relativistic Einstein ring would have formed for different spin $a$ values. The green dashed symmetric circle is the relativistic Einstein ring obtained in the Schwarzschild case where $a=0$.}
\end{figure}

In the Kerr metric, the spherical symmetry is broken by rotation. As a consequence, a perfectly symmetric circular Einstein ring is not formed. Instead, frame-dragging induces an azimuthal dependence on the deflection, resulting in a closed but deformed curve \cite{iyer_strong_2009}. Because our analysis is confined to the equatorial plane, this full deformed ring structure is not visible. Instead, we observe two point images on opposite sides of the black hole, which correspond to the antipodal points of the full, distorted ring.

This effect is explicitly shown in Fig. \ref{98}, which plots the positions of these point images for spin values in the range $-0.5<a<0.5$. The top panels show the first-order images, while the bottom panels show the second-order images. The trend is unambiguous; images associated with retrograde spin ($a<0$) are displaced outward, while those with prograde spin ($a>0$) are displaced inward relative to the Schwarzschild ring position.

This observational signature is a direct and powerful consequence of frame-dragging. The prograde side of the black hole's photon sphere is contracted, pulling the relativistic images closer to the optical axis. The retrograde side is expanded, pushing the images outwards. This results in a characteristic ``displaced and distorted'' ring structure, where the images on one side are significantly closer together than those on the other. Measuring this asymmetry in an observed relativistic ring would provide a direct and quantitative measure of the black hole's spin vector.

\begin{figure}[ht]
  \centering
  \includegraphics[width=0.6\textwidth]{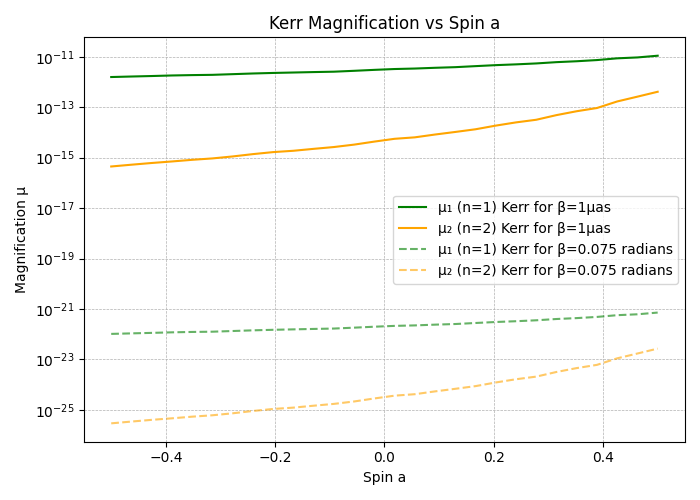}
  \caption{\label{111}Magnifications $\mu_n$ for the first and second order images for the source with angular position $\beta=1\mu$as and $\beta=0.075$rad.}
\end{figure}

The magnifications $\mu_n$ of the first- and second-order relativistic images are shown in Fig. \ref{111} for two source positions: a nearly aligned case ($\beta=1 \mu as$) and a significantly misaligned case ($\beta=0.075 rad$).
A key finding is that the magnifications exhibit a consistent increase with increasing prograde spin, a trend that aligns with the behavior observed for charged black holes in Section \ref{Reissner-Nordstrom Spacetime}. As $a$ increases from $0$ (Schwarzschild) to $0.5$ (extremal), the magnifications for both images increase. This trend is most pronounced for the second-order image in the misaligned case, where the magnification increases by nearly an order of magnitude. This reinforces the counter-intuitive principle established earlier: a modification to the black hole (charge or spin) that weakens the overall deflection can nonetheless enhance the magnification due to a sharper gradient in the lens mapping near the photon sphere.

Furthermore, the results confirm a lensing behavior; the magnification is exquisitely sensitive to the source alignment. For the near-aligned configuration, the magnifications are larger by more than ten orders of magnitude compared to the misaligned case. This underscores that the brightest relativistic images will only be observable when a source, lens, and observer are very closely aligned.

\begin{table}[ht]
\centering
\begin{tabular}{|c|c|c|c|c|c|c|c|c|c|}
    \hline
    \textbf{a} & \textbf{\begin{math}\theta_1 (\mu as)\end{math}} & \textbf{\begin{math}\theta_\infty (\mu as)\end{math}} & \textbf{\begin{math}s (\mu as)=\theta_1 - \theta_\infty\end{math}} & \textbf{$r$} & \textbf{$r_{mag}$} \\
    \hline  -0.5 & 22.674 & 22.672 & 0.002 & 3535.436 & 8.8711\\
    \hline  -0.4 & 21.581 & 21.573 & 0.008 & 2575.970 & 8.5273\\
    \hline  -0.3 & 20.459 & 20.449 & 0.01 & 1853.392 & 8.1601\\
    \hline  -0.2 & 19.302 & 19.289 & 0.013 & 1282.090 & 7.7609\\
    \hline  -0.1 & 18.100 & 18.085 & 0.015 & 846.927 & 7.3196\\
    \hline  0 & 16.848 & 16.826 & 0.022 & 535.491 & 6.8218\\
    \hline  0.1 & 15.521 & 15.488 & 0.033 & 320.436 & 6.2586\\
    \hline  0.2 & 14.095 & 14.044 & 0.051 & 172.457 & 5.5825\\
    \hline  0.3 & 12.519 & 12.427 & 0.092 & 79.236 & 4.7473\\
    \hline  0.4 & 10.692 & 10.482 & 0.21 & 26.467 & 3.5549\\
    \hline  0.5 & - & 6.476 & - & - & -\\
    \hline 
    \end{tabular}
    \caption{\label{tab:5.10}
    Numerical calculation of the remaining inner packed images $\theta_\infty$ which represents the asymptotic position approached by a set of images in the limit $ n\rightarrow \infty$, the separation $s$ between the first relativistic image and the packed images, $r$ the flux of the relativistic image and $r_{mag}$ which is the difference between the magnitude of the first image and all other images for different charge values $a$.}
\end{table}

The flux ratio and image separation, which are critical observables for distinguishing black hole properties, are detailed in Tab. \ref{tab:5.10} and show a dramatic evolution with spin.

As the spin increases from the retrograde limit ($a=-0.5$) to the prograde limit, the angular separation between the first image and the asymptotic pile-up of images, increases by two orders of magnitude from a mere $0.002 \mu as$ to $0.210\mu as$. This significant widening indicates that the set of relativistic images become more ``spread out" for prograde spin, a direct consequence of the increasing strong-field coefficient $\bar{a}$ (see Tab. \ref{tab:9}), which governs the rate of divergence of the deflection angle.

Concurrently, the flux ratio $r$ measures the dominance of the outermost image; a high value means the first image is vastly brighter than the combined inner images. The sharp decrease with prograde spin means that for a rapidly spinning prograde black hole, the outermost image is only about $26$ times brighter than the inner packed images, whereas for a retrograde spinner, the first image dominates by a factor of thousands. The corresponding magnitude difference $r_{mag}$ confirms this trend, decreasing from $8.87$ to $3.55$.

This provides a critical observational signature, the relativistic images of a rapidly spinning prograde black hole will appear as a tighter, more uniform cluster of similar brightness, while those of a retrograde black hole will be dominated by a single, much brighter outermost image with a very faint, closely packed inner set. This clear distinction in the relative brightness and spacing of images offers a promising pathway to measure not only a black hole's spin but also its direction.

The time delay for images on the same side can be found \cite{bozza_time_2004}:
\begin{equation}\label{eq1}
    \triangle T^s_{n,l}=2\pi(n-l)\frac{\Tilde{a}}{\Bar{a}}+2\sqrt{\frac{B_pJ_p}{A_p}}\left[e^\frac{\Bar{b}-2l\pi\pm\beta}{2\Bar{a}}-e^\frac{\Bar{b}-2n\pi \pm\beta}{2\Bar{a}}\right]\approx 2\pi(n-l) J_p\,,
\end{equation}
and the time delay for images on the opposite side of the lens is \cite{bozza_time_2004}:
\begin{equation}
    \triangle T_{n,l}^o = [2\pi n +\beta - \Bar{b}(a)] \frac{\Tilde{a}(a)}{\Bar{a}(a)} +\Tilde{b}(a) -  [2\pi l -\beta - \Bar{b}(-a)] \frac{\Tilde{a}(-a)}{\Bar{a}(-a)} +\Tilde{b}(-a)\,.
\end{equation}
An approximate expression has been found by \cite{hsieh_gravitational_2021} such that: 
\begin{equation}
    \triangle T_{n,l}^o = 3\sqrt{3} (n-l)\pi + 2 a [-2 (1+n+l) \pi -3 \sqrt{3} \ln(3(7- 4\sqrt{3}))]\,.
\end{equation}

\begin{table}[ht]
    \centering
    \begin{tabular}{|c|c|c|c|c|c|c|c|c|c|c|c|c|}
        \hline
        a & -0.5  &-0.4  &-0.3   &-0.2  &-0.1  &0 &0.1  &0.2  &0.3  &0.4 & 0.5 \\
        \hline
        $\triangle T^s_{2,1}$ (min)   & 10.09   &9.60     &9.10    &8.58 &8.05   &7.49   &6.89    &6.25    &5.53    &4.66   &2.88 \\ 
        \hline
        $\triangle T^o_{1,1}$ (min) & 4.98 &3.99 &2.99 &1.99 &0.99 &0 &-0.99 &-1.99 &-2.99 &-3.99 &-4.98\\ \hline
    \end{tabular}
    \caption{Time delay calculations for the first and second order images on the same $\triangle T^s_{2,1}$ (min) side of the lens. Time delays for the first-order image on opposite sides of the lens $\triangle T^o_{1,1}$ (min) due to the prograde and retrograde spin calculated for $\beta=0$. The negative sign indicates that direct photons (co-rotating with the black hole spin) reach the observer faster than retrograde (counter-rotating) photons. $a=0$ corresponds to the Schwarzschild BH.}
    \label{tab:22}
\end{table}

Tab. \ref{tab:22} 
The time delay between images on the same side, $\Delta T^s_{2,1}$, exhibits a monotonic decrease with increasing prograde spin, falling from $10.09$ minutes at $a=-0.5$ to $2.88$ minutes at $a=0.5$. This trend is a direct consequence of the spin-dependent contraction of the photon sphere. As established previously, the critical impact parameter $J_p$ decreases significantly for prograde orbits. Since the dominant term in Eq. \ref{eq1} is proportional to $J_p$, the time it takes for successive windings around the black hole is reduced, leading to a shorter delay between higher-order images.

A more profound signature of rotation is found in the delay between images on opposite sides of the black hole, $\Delta T^o_{1,1}$. This delay is not merely a function of winding number but of the photon's orbital direction relative to the spin. For retrograde spin ($a<0$), $\Delta T^o_{1,1}$ is positive, meaning that the retrograde photon arrives after the prograde one. As the spin progresses ($a>0$), this delay becomes negative, reaching $-4.98$ minutes at $a=0.5$.

This sign change is a direct manifestation of frame-dragging. A prograde (co-rotating) photon is effectively ``carried along" by the spinning spacetime, traversing its null geodesic in a shorter coordinate time compared to a retrograde (counter-rotating) photon, which must fight against the dragged spacetime. Therefore, the image on the prograde side of the black hole can arrive at the observer before the image on the retrograde side, a unique relativistic effect that provides an unambiguous method to determine the direction of a black hole's spin.

\section{Discussion and Conclusion}

This study has provided a comprehensive analysis of how black hole charge (in the RN metric) and spin (in the Kerr metric) influence strong gravitational lensing observables, with specific application to the supermassive black hole Sagittarius A.

Our investigation reveals a fundamental trend; both electric charge and prograde spin act to weaken the effective strength of the gravitational lens in the strong-field regime. This is demonstrated by the systematic decrease in the radius of the photon sphere ($x_p$), the critical impact parameter ($J_p$) and the angular positions of the relativistic images ($\theta_1, \theta_2$) as the charge $q$ or the prograde spin $a$ increases. The physical origin of this weakening differs, charge introduces a repulsive gravitational component, while spin contracts the photon sphere via frame-dragging, but the overarching effect on the lensing scale is consistent.

However, a more nuanced picture emerges from the detailed observables. We identified several unique signatures that can distinguish between these black hole properties:
\begin{enumerate}
    \item The Magnification Problem: Despite producing weaker deflection, both charge and prograde spin lead to a mild increase in the magnification of relativistic images. This highlights that the brightness of an image is not solely determined by the total bending of light, but by the gradient of the deflection angle, which becomes sharper near the photon sphere of a charged or spinning black hole.

    \item The Asymmetric Kerr Signature: The spin of a black hole breaks spherical symmetry, leading to unambiguous observational consequences. Relativistic images are displaced, with prograde images shifting inward and retrograde images shifting outward. Most strikingly, the time delay between images on opposite sides of the lens, $\Delta T^o_{1,1}$, changes sign. For a prograde-spinning black hole, the image on the approaching side can arrive before the image on the receding side, a direct and measurable effect of frame-dragging.

    \item Distinct Image Architectures: The flux ratio $r$ and image separation $s$ evolve differently. A highly charged or prograde-spinning black hole produces a set of relativistic images that are more tightly packed and have more uniform brightness. In contrast, a Schwarzschild and retrograde-spinning black hole produce a dominant, bright outermost image with a faint, closely packed inner series.
\end{enumerate}


In conclusion, while charge and prograde spin both contract the scale of strong lensing, their distinct imprints on image magnification, asymmetry, and timing provide a rich set of observables. Future work should integrate these results with realistic astrophysical environments and assess their detectability with the next generation of very-long-baseline interferometry, bringing us closer to a complete characterization of the compact objects that dominate our galactic center.

\bibliographystyle{unsrt}
\bibliography{references}

\end{document}